\documentclass[12pt]{iopart}
\usepackage{iopams}
\usepackage{epsfig}
\usepackage{color}

\newcommand{\bea}{\begin{eqnarray}}
\newcommand{\eea}{\end{eqnarray}}

\newcommand{\vect}[1]{\mathbf{#1}}

\newcommand{\di}{\displaystyle}

\newcommand{\kbt}{k_{\rm B}T}
\newcommand{\mc}{\mathcal}

\begin{document}

\title{Stability and interactions
of nanocolloids at fluid interfaces: effects of capillary waves and line tensions$\;$}

\author{H.~Lehle\dag and M.~Oettel\ddag}

\address{\dag\
Max--Planck--Institut f\"ur Metallforschung, Heisenbergstr. 3, D-70569 Stuttgart, \\
Institut f\"ur Theoretische und Angewandte Physik, Universit\"at Stuttgart,
 Pfaffenwaldring 57, D-70569 Stuttgart, Germany}
\address{\ddag\
Johannes--Gutenberg--Universit\"at Mainz, Institut f\"ur Physik (WA 331), 
Staudinger Weg 7, D-55128 Mainz, Germany}

\begin{abstract}
  We analyze the effective potential for nanoparticles trapped at a
  fluid interface within a simple model which incorporates surface and
  line tensions as well as a {thermal} average over interface fluctuations
  (capillary waves). For a single colloid, a reduced  steepness of the potential
  well hindering movements out of the interface plane  compared
  to rigid interface models is observed, and an instability of the 
  capillary wave 
  partition sum in case of negative line tensions is pointed out.     
  For two colloids, averaging over the capillary waves leads to an effective
  Casimir--type interaction which is long--ranged, {power-like in the
  inverse distance} but {whose power} sensitively depends
  on possible restrictions of the {colloid degress of freedom}. 
  A nonzero line tension
  leads to changes in the magnitude but not in the functional form
  of the effective potential asymptotics.    
\end{abstract}

\pacs{24.60.Ky,68.03.Kn,82.70.Dd}

\submitto{\JPCM}

\section{Introduction}

The effective forces between rigid objects immersed
in a fluctuating medium have attracted a steadily growing interest
because their understanding allows one to design and tune them by choosing
suitable media and boundary conditions and by varying the thermodynamic state
of the medium. Possible applications range from micromechanical systems to
colloidal suspensions and embedded biological macromolecules.
Accordingly, these fluctuations may be the
 zero--temperature, long--ranged quantum fluctuations of the
electromagnetic fields giving rise to the
original Casimir effect \cite{Bor01} between flat or corrugated immersed
metallic bodies \cite{Jaf04, Bue04}.
Other examples for fluctuation induced long--ranged effective forces between
immersed objects involve media such as
bulk fluids near their critical point \cite{Kre94,Bec07}, 
membranes \cite{Kar99} or interfaces \cite{Kai05}.

A quasi two--dimensional realization of a fluctuating system with
long--ranged correlations is given by  
a liquid--liquid or liquid--vapor interface
{along with its capillary wave excitations}, for an experimental
study in real space see Ref.~\cite{Aar04}. Consequently rigid objects such
as colloids which are trapped at the interface are subject to an effective
force generated by the fluctuating capillary waves. Theoretical
investigations of this phenomenon have dealt with the effective
interaction between pointlike \cite{Kai05} or rodlike \cite{Gol96} objects.
If the particles are held fixed, and the interface is pinned at their 
surfaces, the capillary wave mediated interaction between the particles is
a direct analogon of the original Casimir effect for a 
two--dimensional Gaussian scalar field
with Dirichlet boundary conditions at the one--dimensional
boundaries, given by the three--phase contact lines
(for a general treatment on the Casimir effect in scalar fields, see
Ref.~\cite{Emi08}). However, the interface
and the colloids are embedded in three--dimensional space, and therefore
the colloids and the position of the three--phase contact lines
may fluctuate. It has been shown in Refs.~\cite{Gol00,Leh06,Leh07} 
that this situation 
corresponds theoretically to a Casimir problem with {\em fluctuating} 
boundary conditions. Depending on the type of admitted boundary fluctuations,
the asymptotics of the resulting Casimir interaction varies considerably.
This is an effect absent in the case of objects immersed in 
three--dimensional bulk systems.  
Moreover, the fluctuations of the colloids in direction perpendicular to the
interface plane influence their stability, i.e. the activation energy to 
desorb them from the interface. This aspect has received little attention
so far, as often rigid--interface models are employed to discuss the
energetics of trapped colloids at interfaces (see, e.g., 
Refs.~\cite{Pie80,Ave96}).
     
{In this work, both questions, the stability of single colloids and their
Casimir--like interactions in the presence of fluctuating capillary waves, are analyzed
with particular attention to the effects of a nonzero line tension.}
The paper is structured as follows: In Sec.~\ref{sec:single} we discuss 
a free energy model for the entrapment and stability 
of a single colloid at a fluid
interface. In order to be self--contained, the predictions of the rigid
interface models incorporating surface and line tensions are summarized
in Subsecs.~\ref{sec:surface} and \ref{sec:line}, whereas the effect
of capillary waves is treated in Subsec.~\ref{sec:cw}. 
In Sec.~\ref{sec:two} we discuss the fluctuation--induced interaction 
between two trapped colloids, making use of the fluctuation Hamiltonians derived
for the single colloid problem. The importance of the type of admitted
boundary fluctuations on the effective interaction is highlighted 
with the exemplary discussion
of two cases: (i) both colloids fluctuate freely and (ii) the colloids
are held fixed. For both cases, line tensions modify the magnitude of the
effective interaction but are shown not to lead to qualitatively new       
behavior.

\section{Free energy model for spherical colloids trapped at interfaces}
\label{sec:single}

In this section, we consider the free energy of a single, spherical colloid 
with radius $R$ trapped
at an interface between two phases I and II which is assumed to be flat in 
equilibrium. 
The height of the center of the colloid above the interface 
is denoted by $z$, and we look for the free energy $F$ of the colloid with $z$ 
fixed, i.e. $F\equiv F(z)$ describes a constrained free energy and the equilibrium
free energy for the colloid follows upon minimization with respect to $z$.
This constrained free energy can be viewed as an effective potential, in which
the colloid fluctuates.  
 In Subsecs.~\ref{sec:surface} and \ref{sec:line} we introduce
the surface and line tension contributions to $F(z)$ under the assumption that
the interface stays rigidly flat. In Subsec.~\ref{sec:cw} the effect of 
interface fluctuations on the constrained free energy $F(z)$ is calculated
in a perturbative manner, employing the capillary wave concept. 

\subsection{Surface tension}
\label{sec:surface}

\begin{figure}
 \begin{center}
  \epsfig{file=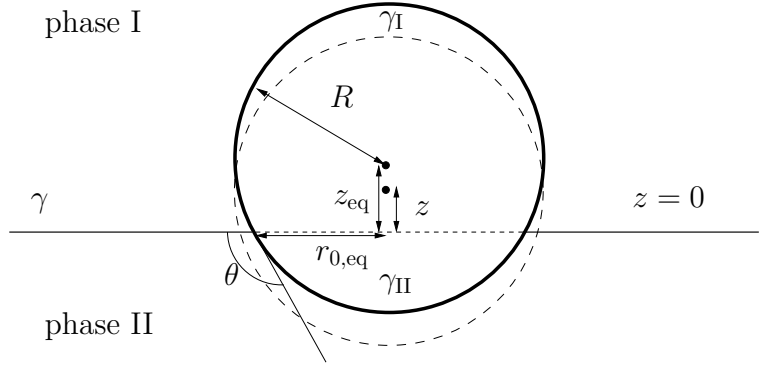, width=10cm}
 \end{center}
 \caption{Side view of a single colloid with radius $R$ trapped at an 
interface ($z=0$).
The full circle shows the colloid in equilibrium, with ist center located
at a height $z_{\rm eq}$ above the interface. The radius of the circular
three--phase contact line is given by $r_{0,{\rm eq}}$. The dashed circle 
shows the colloid out of equilibrium at a center position $z$. The surface 
tensions phase I/phase II, colloid/phase I, and colloid/phase II are denoted
by $\gamma$, $\gamma_{\rm I}$, and $\gamma_{\rm II}$, respectively.
The physical contact angle $\theta$ differs from Young's angle
$\cos\theta_0=(\gamma_{\rm I}-\gamma_{\rm II})/\gamma$ for a non--vanishing
line tension. }
 \label{fig:rigid}
\end{figure}

Let $\gamma$ denote the surface tension of the fluid interface and
$\gamma_{\rm I[II]}$ denote the surface tension of the colloid surface (with area
$A_{\rm I[II]}$) exposed to phase I [II] (see Fig.~\ref{fig:rigid}). 
The free energy $F(z)$, measured with respect to the configuration
where the colloid is completely immersed in phase I, is given by \cite{Pie80} 
\bea
   F(z) &=& \gamma_{\rm I}\,(A_{\rm I} -4\pi R^2) + 
   \gamma_{\rm II}\,A_{\rm II} - \gamma\, A \;. 
\eea  
Here, $A$ is the cross section area of the colloid with the interface. 
Introducing the reduced quantities $\hat z = z/R$, 
$\hat F(\hat z)= F(z)/(\pi\gamma R^2)$, and Young's angle $\theta_0$ via 
$\cos\theta_0=(\gamma_{\rm I}-\gamma_{\rm II})/\gamma$, the free energy becomes
\bea
 \label{eq:f_pie}
   \hat F(\hat z) &=& (\hat z + \cos\theta_0)^2 -(1+\cos\theta_0)^2 \;.
\eea
According to this model, the free energy (or single colloid effective potential)
is harmonic with spring constant $k= F''(z_{\rm eq})= 2\pi\gamma$. 
The equilibrium position is
given by $\hat z_{\rm eq}= - \cos\theta_0$ and the depth of the harmonic well
defines the activation energy $E_{\rm a} = \pi\gamma R^2 (1-|\cos\theta_0|)^2)$.
Thus we see that for interfaces of simple fluids ($\gamma \sim 10^{-2}$ N/m) 
the activation energy becomes comparable to the thermal energy 
$\beta^{-1}=\kbt$
only for nanoscopic colloids with radius $R \simeq 1$ nm.
At such small scales the application of the simple, thermodynamic concept
of surface tension seems doubtful. 
Nonetheless, recent investigation using
computer simulations \cite{Bre99,Bre99a} have indicated that the behavior 
of truly 
nanoscale particles can be described within a wide range of conditions by such 
a phenomenological
thermodynamic model upon introduction of particle size dependent surface tensions 
and three--phase line tensions.

\subsection{Line tension}
\label{sec:line}

The line tension was introduced by Gibbs~\cite{Gib61,Row02} to define the
excess free energy associated to the line where three phases meet. The accurate
experimental measurement of the line tension has been a considerable challenge.
In fact, the uncertainty in the order of magnitude of the line tension has generated
a considerable number of  studies. The interested 
reader is referred to a recent review devoted to the current status of the 
three--phase line tension~\cite{Ami04}.
On theoretical grounds the line tension  can be either positive or negative and
it is expected to be a small force~\cite{Row02}, $10^{-11}$~N. For simple
fluids away from criticality, dimensional analysis gives $\tau \sim \kbt/\sigma
$ where $\sigma \sim 0.1$ nm is a typical atomic length scale.
 Line tensions inferred from experiments 
span several orders of magnitude, $10^{-12}-10^{-6}$~N~\cite{Ami04,Dre96}, 
reflecting the variety of experimental
techniques (which are all indirect and where systematic errors can hardly be
estimated) and the variety of materials used.
More recent experimental efforts point towards line tensions with
magnitudes at the lower
end of the range given above (see, e.g., the discussion in Ref.~\cite{Schim07}
and references therein).
 It is not clear 
from the outset that the concept of line tension is unique, i.e. independent from 
notional shifts of the interfaces 
(which also shift the location of the contact line) that should be
permitted within the  molecularly diffuse interface region.
This question is treated in detail in Ref.~\cite{Schim07} which lines out a 
protocol how to define and relate line tensions obtained in different experimental
setups.

\begin{figure}
 \begin{center}
   \epsfig{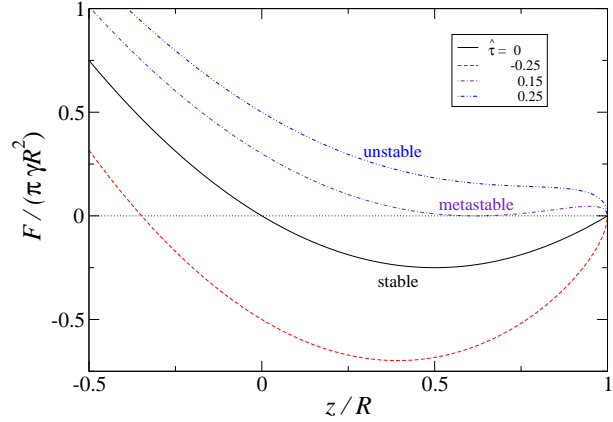}
 \end{center}
 \caption{Plots of the reduced effective potential $\hat F(\hat z)$ for 4 different
 values of the reduced line tension $\hat \tau$. With increasing $\hat \tau$,
 the configuration crosses from the stable via the metastable to the unstable
 regime. Young's angle is set to $\theta_0=120^o$ which corresponds to the
 contact angle of $\mu$m--sized polystyrene colloids at an oil/water interface
 \cite{Par07}.   }
 \label{fig:stability}
\end{figure}

For our purposes, we may assume that a specific definition for the interface
locations has been adopted with regard to which surface tensions are defined
(see Fig.~\ref{fig:rigid}).
If we associate a line tension $\tau$ with the energy of the three--phase
contact line (which is a circle with radius $r_0=\sqrt{R^2-z^2}$), the free energy 
of Eq.~(\ref{eq:f_pie}) must be amended as follows:
\bea
  \label{eq:f_ave}
   \hat F(\hat z) &=& (\hat z + \cos\theta_0)^2 -(1+\cos\theta_0)^2 +
    2\sqrt{1-\hat z^2}\,\hat \tau \;.
\eea
Here, the reduced line tension is given by $\hat \tau =\tau/(\gamma R)$.
The new equilibrium position of the colloid is determined by an equation
of fourth order in $\hat z$:
\bea
 \label{eq:zeq}
  \hat z_{\rm eq}\left( \frac{\hat \tau}{\sqrt{1-\hat z_{\rm eq}^2}} -1\right) &=&
   \cos\theta_0 \;. 
\eea
The spring constant $k= F''(z_{\rm eq})$ of the potential well near the 
minimum is also modified and becomes
\bea
 \label{eq:k1}
 k= 2\pi\gamma (1 - \hat \tau /\hat r_{0,\rm eq}^3)
\eea
where $\hat r_{0,\rm eq} = \sqrt{1-\hat z_{\rm eq}^2}$. 
Note that the effective potential
defined by Eq.~(\ref{eq:f_ave}) is not harmonic anymore. Positive values for
the line tension lead to a shallower well, inducing  metastability for the trapped
colloid \cite{Ave96}. Large positive line tensions lead to the desorption of the
colloid from the interface (see Fig.~\ref{fig:stability} for the variation
of the effective potential $\hat F(\hat z)$ with $\hat \tau$ for $\theta_0=120^o$). 
Thus, within this simple model, a first order transition between 
colloids desorbed in one of the bulk phases and adsorption to the interface is 
possible \cite{Ave96}.

\subsection{Capillary waves}
\label{sec:cw}

In thermal equilibrium, the interface is not sharp but acquires a finite
thickness through density fluctuations. 
In a coarse--grained picture, these density fluctuations correspond to
fluctuations of the mean interface position $u(\vect x)$ around the equilibrium
position $u=0$. Here, $\vect x = (x,y)$ is a vector in the equilibrium
interface plane $z=0$. 
For wavelengths of these interface position 
fluctuations which are larger than the correlation length
in the bulk phases, the free energy of an interface configuration $u(\vect x)$ 
is given by the surface energy of this configuration:
\bea
 \label{eq:f_cw1}
  F_{\rm men} = \gamma \int_{S_{\rm men}} d^2x \left[\sqrt{1+ (\nabla u)^2}
    + \frac{u^2}{\lambda_c^2}\right]\;.
\eea    
Here, $S_{\rm men}$ is the interface or meniscus area projected onto
the plane $z=0$ of the reference interface, i.e. it is the whole plane
$z=0$ with colloids which are possibly trapped at the interface ``cut out". 
The free energy in Eq.~(\ref{eq:f_cw1}) contains an additional
term which accounts for the costs in gravitational energy
associated with the meniscus fluctuations. It involves the capillary length
$\lambda_c$ given by
$\lambda_c=[\gamma/(|\rho_{\rm II}-\rho_{\rm I}|g)]^{1/2}$,
where
$g$ is the gravitational constant and
$\rho_i$ the mass density in phase $i$.
Usually, in simple fluids $\lambda_c$ is in the range
of millimeters. Surface excitations whose free energy are described by
Eq.~(\ref{eq:f_cw1}) are termed {\em capillary waves}, and their
effect on static and dynamic properties of liquid interfaces
is a subject of lively interest 
\cite{Mec99,Dai00,Mil02,Dai03,Tol04,Vin05,Tar07}.  

The effective potential $F(z)$ for the colloid whose center
is fixed at height $z$ above the equilibrium interface can be obtained
through the partition function of the capillary waves:
\bea
  F(z) & = & - \kbt \ln \mc{Z} \;, \\
 \label{eq:Zsingledef}
  \mc{Z} &=&  \mc{Z}_0^{-1} \int \mathcal{D}u \,\exp(-\beta\mc{H})
  \qquad \qquad (\beta^{-1}=\kbt)\;.
\eea  
Here, $\mc{Z}_0$ is a suitable normalization factor. The Hamiltonian $\mc{H}$
which enters the Boltzmann weight for a certain interface configuration
$u(\vect x)$ is the difference in free energy between the configuration
$\{u(\vect x),z\}$ (describing the interface and colloid position) and the
equilibrium configuration $\{u=0,z_{\rm eq}\}$ which we call the reference
configuration. The equilibrium position of the colloid is determined
through Eq.~(\ref{eq:zeq}). Therefore
\bea
 \label{eq:hs}
 \fl
  \mc{H}[u(\vect x),z] &=& 
     (F_{\rm men}[\{u(\vect x),z\}] - F_{\rm men}[\{0,z_{\rm eq}\}])
             +\gamma_{\rm I}\Delta A_{\rm I}
             +\gamma_{\rm II}\Delta A_{\rm II} + \tau \Delta L 
\eea  
The difference in the interface areas colloid/phase I and colloid/phase II
between the configuration $\{u,z\}$ and the reference configuration is 
given by $\Delta A_{\rm I}$ and $\Delta A_{\rm II}$, respectively, and the
difference of the three--phase contact line length between these 
configurations is given by $\Delta L$. In general, $\mc{H}[u(\vect x),z] $
is a complicated functional of $u$ and $z$. In order to reduce it to a 
tractable expression which allows the analytical determination of the
functional integral in Eq.~(\ref{eq:Zsingledef}), we perform a 
Taylor expansion in $u$ and $z$ \cite{Oet05}. 
To quadratic order, this implies that the Hamiltonian can be split into
a two--dimensional ``bulk" term $\mc{H}_{\rm cw}$ and a one--dimensional
``boundary" term $\mc{H}_{\rm b}$:
\bea
 \label{eq:hs_split}
  \mc{H}[u(\vect x),z] &=& \left.\mc{H}_{\rm cw}[u(\vect x)]
    \right|_{\vect x \in S_{\rm men,ref}} +
   \left.\mc{H}_{\rm b}[u(\vect x),z] 
   \right|_{\vect x \in \partial S_{\rm men,ref}}\;.
\eea
The two--dimensional ``bulk" area $S_{\rm men,ref}$ is given by the 
interface in the reference configuration (i.e. the plane $z=0$ with the 
colloid cut out), and the boundary line $\partial S_{\rm men,ref}$ is
given by the three--phase contact line in the reference configuration.
With these definitions $\mc{H}_{\rm cw}$ corresponds to the usual 
capillary wave Hamiltonian {\cite{Buf65}:}
\bea
 \label{eq:hcw}
 \mc{H}_{\rm cw}[u(\vect x)] &=& \frac{\gamma}{2} \int_{S_{\rm men,ref}} d^2x\,
\left[ (\nabla u)^2
+\frac{u^2}{\lambda_c^2}
\right] \;.
\eea  
The boundary term $\mc{H}_{\rm b}$ only depends on the difference
$h=z-z_{\rm eq}$ and {the} vertical
position of the contact line $f$. The latter is expressed by its Fourier
transform:
\bea
\label{eq:conlin}
f(\varphi) =
u(\partial S_{\rm men,ref})=\sum_{m=-\infty}^\infty  P_{m}\,e^{{\rm i}m\varphi}
\;.
\eea
In Eq.~(\ref{eq:conlin}), the polar angle $\varphi$ is defined on the reference
contact line circle $\partial S_{\rm men,ref}$.
The Fourier coefficients $P_{m}$ are referred to as
contact line multipoles below, and 
since the contact line height $f(\varphi)$ is real, 
$P_m=P^*_{-m}$ holds. With these definitions, the boundary term 
acquires the form
\bea
 \label{eq:hb}
  \mc{H}_{\rm b}[u(\vect x),h] &=& \mc{H}_{\rm b,1} + \mc{H}_{\rm b,2}\,, \\
 \label{eq:hb1}
  \mc{H}_{\rm b,1} &=& \frac{\pi\gamma}{2}\left[ 2(P_{0}-h)^2+ 
   4\sum_{m=1}^{\infty} |P_{m}|^2\right]\;, \\
 \label{eq:hb2}
  \mc{H}_{\rm b,2} &=& \frac{\pi \gamma}{2} \frac{R^3}{r_{0,\rm eq}^3}\,
   \hat \tau
       \left[ -2 \, (P_{0}-h)^2
       + 4 \sum_{m=1}^{\infty} \left( m^2-1 \right) \,|P_{m}|^2 \right] \;.
\eea 
As before, $r_{0,\rm eq}=\sqrt{R^2-z_{\rm eq}^2}$ is the radius
of the circle enclosed by the reference contact line and
$\hat \tau =\tau/(\gamma R)$ is the reduced line tension.
The detailed derivation of the two terms contributing to $\mc{H}_{\rm b}$
can be found in \ref{app:boundary}. Note that $\mc{H}_{\rm b,1}$ which
describes the change in colloid surface energy upon shifting the contact line 
is strictly positive definite, whereas $\mc{H}_{\rm b,2}$ is not
positive definite, regardless of the sign of $\tau$.

The partition function $\mc{Z}$ (Eq.~(\ref{eq:Zsingledef})) can be written 
such that the integral
over contact line fluctuations $\int\mc{D}f$ appears explicitly:
\bea
 \label{eq:Z1single} \fl
\mathcal{Z}=\mathcal{Z}_0^{-1} \int \mathcal{D}u\,
\exp \left(-\beta\mathcal{H}_{\rm cw}[u,z]  \right)\;
\int \mathcal{D}f \!\!\!
\prod_{{\bf x} \in \partial S_{\rm men,ref}} 
\delta [u({\bf x})-f({\bf x})]
\exp \left(-\beta\mathcal{H}_{{\rm b}}[f,h] \right)
\;. \quad
\eea
The integration measure for the contact line fluctuations
is given by $\mc{D}f=dP_0\prod_{m>0} d{\rm Re}P_{m}d{\rm Im}P_{-m}$.
However, in this form the 2d-``bulk" fluctuations $u$ are not yet
separated from the ``boundary" fluctuations $f$. This can be achieved
by splitting the field $u$ of the local interface position into
a mean--field and a fluctuation part, $u=u_{\rm mf}+v$. The
mean--field part solves the Euler--Lagrange equation
$ (-\Delta+\lambda_c^{-2})\,u_{\rm mf}=0$ with the boundary condition
$u_{\rm mf}\,|_{ \partial S_{\rm men, ref}}=f $. 
Consequently the fluctuation part
vanishes at the contact line:
$v\,|_{ \partial S_{\rm men, ref}}=0$.
Then the partition function  
$\mathcal{Z}=\mathcal{Z}_{\rm fluc}\mathcal{Z}_{\rm mf}$
factorizes into a product of a fluctuation part
independent of
the boundary conditions
 and a  mean field part which depends on the fluctuating contact line 
$f$:
\begin{eqnarray}\label{eq:Zsepsingle}
 \mathcal{Z}_{\rm fluc} &=& \mathcal{Z}_0^{-1} \int \mathcal{D}v\,
 \prod_{{\bf x} \in \partial S_{\rm men, ref}}\delta ( v({\bf x}))
\exp \left(-\beta\mathcal{H}_{\rm cw}[v] \right)\;,   \\
 \label{eq:Zmf_single}
 \mathcal{Z}_{\rm mf} (h) &=& 
\int \mathcal{D}f
\, \exp\left\{ -\frac{\beta\gamma }{2} 
\oint_{\partial S_{\rm men, ref}} \!\!\!\!\!\!\!\!\!\!
  d\ell\, f({\bf x})\,( \partial_n u_{\rm mf}({\bf x}))\right\}
\exp \left(-\beta\mathcal{H}_{{\rm b}}[f,h] \right)
 \;. \nonumber
\end{eqnarray}
The first exponential in $\mathcal{Z}_{\rm mf}$ stems from applying Gauss' theorem
to the energy associated with $u_{\rm mf}$. In this term
$ \partial_n u_{\rm mf}$ denotes the normal derivative of the mean--field
solution towards the {interior} of the circle $\partial S_{\rm men, ref}$, 
and $d\ell$ is
the infinitesimal line segment on $\partial S_{\rm men, ref}$.

Since $\mathcal{Z}_{\rm fluc} $ does not depend on the colloid position
$h=z-z_{\rm eq}$, it only contributes an additive constant to the 
effective potential. In $\mathcal{Z}_{\rm mf}$, only the monopole
fluctuations of the contact line $P_0$  are coupled to $h$ 
(see Eqs.~(\ref{eq:hb1}) and (\ref{eq:hb2})). The mean--field energy term
(the first exponential in $\mathcal{Z}_{\rm mf}$, Eq.~(\ref{eq:Zmf_single})) 
is diagonal in the multipole moments $P_m$ (see \ref{app:umf}):
\bea
 \label{eq:umf}
 -\frac{\beta\gamma }{2}
\oint_{\partial S_{\rm men, ref}} \!\!\!\!\!\!\!\!\!\!
  d\ell\, f({\bf x})\,( \partial_n u_{\rm mf}({\bf x}))
 \stackrel{\lambda_c \gg R}{\simeq} 
  -\pi\beta\gamma \left( \frac{P_0^2}{\ln \hat \lambda_c} + 2 
   \sum_{m> 0} m|P_m|^2 \right)\;, 
\eea
with the reduced capillary length given by 
$\hat \lambda_c = 1.12\lambda_c/r_{0,{\rm eq}}$.
Therefore only the integral over the monopole fluctuation $P_0$
yields a dependence on $h$ in the partition function:
\bea
 \fl
  \mc{Z} &=& \mc{Z}'_0[v,P_{m\,(m\not =0)}] \,
  \int dP_0 
   \exp\left\{ - \pi \beta\gamma \left[ 
   \frac{P_0^2}{\ln\hat\lambda_c} + (P_0-h)^2\, 
   \left(1 - \frac{\hat \tau}{\hat r_{0,{\rm eq}}^3}\right) 
   \right] \right\} \\
 \fl
        &=& \mc{Z}'_0[v,P_{m\,(m\not =0)}]\,
  \left[ \beta\gamma\left( \frac{1}{\ln\hat\lambda_c} +
    1 - \frac{\hat \tau}{\hat r_{0,{\rm eq}}^3}\right) \right]^{-1/2}\;
  \times \\
  \fl && \qquad \qquad
  \exp\left\{ - \pi \beta\gamma\, h^2\,  
   \frac{ 1 - \frac{\di \hat \tau}{\di \hat r_{0,{\rm eq}}^3}}
   {1+\ln\hat\lambda_c
   \left(1-\frac{\di \hat \tau}{\di \hat r_{0,{\rm eq}}^3}\right)}
   \right\}
\eea 
The $h$--independent contributions have been put into the new normalization
factor $\mc{Z}'_0$.
Thus the effective potential for the colloid moving around its equilibrium
position is given by
\bea
   F(z) & = & - \beta^{-1} \; \ln \mc{Z} \\
 \label{eq:f_single_cw}
        & = & {\rm const.} + \pi \gamma\, (z-z_{\rm eq})^2\,
   \frac{ 1 - \frac{\di \hat \tau}{\di \hat r_{0,{\rm eq}}^3}}
   {1+\ln\hat\lambda_c
   \left(1-\frac{\di \hat \tau}{\di \hat r_{0,{\rm eq}}^3}\right)}
   \;.
\eea
The spring constant $k=F''(z_{\rm eq})$ of the effective potential
can be compared to the one derived for the case of a rigid interface
(Eq.~(\ref{eq:k1}). It is seen that we recover the latter upon neglecting
the term involving the capillary length $\lambda_c$. However, for 
a physical situation, the capillary length is not small and greatly
diminishes the steepness of the potential well (for
colloids with $R=10$ nm at an air--water interface, $\theta_0 \simeq
90^o$ and negligible line tensions, $k$ is reduced by a factor of 14).
In accordance with the Goldstone boson character of capillary waves,
the spring constant vanishes for $\lambda_c \to \infty$ since the whole
interface can move with no energy cost upon shifting the colloid.  

In accordance with the rigid interface result
(Eq.~(\ref{eq:k1})), the effective potential becomes unstable
for $\hat \tau > \hat r_{0,{\rm eq}}^3$.
However, this perturbative calculation of the effective potential $F(z)$ allows
no conclusion on the height of the energy barrier, which exists
in the metastable regime of positive line tensions 
(see Fig.~\ref{fig:stability}). One may speculate that the considerable
reduction of the potential well steepness by capillary waves goes along 
with a reduction of the barrier height and would thus facilitate
particle desorption.\footnote{This effect might be expected quite
in similarity to the effective reduction of the barrier height in a  
double-well potential
for a single quantum--mechanical particle \cite{Klei_book}.}
  
For negative line tensions, the shape of the effective potential in
Eq.~(\ref{eq:f_single_cw}) is hardly affected since $\ln\hat\lambda_c \gg 1$.
However, an instability shows up in the partition function which is 
contained in the factor $\mc{Z}_0'$. This factor contains
a contribution of the form
\bea
 \label{eq:instability}
  \mc{Z}_0' & \propto & \mc{D}'f 
    \exp\left\{ - 2\pi\beta\gamma \, \sum_{m>0} |P_m|^2
    \left[ (m+1) + (m^2-1)\frac{\di \hat \tau}{\di \hat r_{0,{\rm eq}}^3}  
   \right] \right\}
\eea
where the measure for the contact line fluctuations does not contain the
monopoles, $\mc{D}'f = \int \prod_{m>0} d{\rm Re}P_{m}d{\rm Im}P_{-m}$. 
Clearly, for arbitrarily small but negative $\tau$ there exists a
critical multipole order $m_c$ above which the exponent becomes
positive and thus the partition function becomes infinite. Taken at face
value, for negative line tensions the interface would become unstable
by forming ripples with small wavelengths near the colloid. In a physical
situation, there is however a lower cutoff in the wavelength of these
ripples set by the colloid surface roughness which presumably adds
a positive energy penalty to higher multipole fluctuations of the
contact line.  

As is well--known, the capillary length $\lambda_c$ serves effectively as 
an infrared cutoff for the capillary wave spectrum. Although the capillary
wave model used in this work was derived assuming gravitational
damping of the capillary waves (see Eq.~(\ref{eq:f_cw1})), the dependence
of all resulting expressions on $\lambda_c$ is not specific to that
form of the capillary wave Hamiltonian \cite{Oet05}. Such an infrared cutoff 
results equally well from a finite system size (e.g. a two--phase system
with trapped colloids in a container with extension $L$ 
or colloids trapped on a droplet with radius $R_{\rm d}$).
Thus for such systems, the effective potential (Eq.~(\ref{eq:f_single_cw}))
is obtained by just replacing $\lambda_c$ by $L$ or $R_{\rm d}$.
Since for smaller system sizes the influence of the line tension on $F(z)$
is more pronounced, it is conceivable to obtain a value
for $\tau$ from the fluctuations in the colloid position which sample
$F(z)$. As can be seen from Eq.~(\ref{eq:f_single_cw}), no knowledge
of the surface tension $\gamma_{\rm I}$ and $\gamma_{\rm II}$
(or Young's angle $\theta_0$) is required, but the modified
contact angle $|\sin\theta|=\hat r_{0,\rm eq}$ enters.   
Certainly,  
an experimental realization appears to be difficult
because of the difficulty in determining $\theta$,
and one could resort to simulations in a first step which determine the
colloid fluctuations with varying system size and in which $\theta$ can be
determined straightforwardly.

\section{Fluctuation induced forces between two colloids}
\label{sec:two}

The previous considerations can be extended to the case of two colloids which 
are trapped at the interface at distance $d$. Clearly, if both colloids are at 
their equilibrium position (defined by Eq.~(\ref{eq:zeq})) and capillary waves 
are neglected, the interface is flat and therefore no interface--mediated 
interactions are present. If, by some external force, the colloids
are moved away from equilibrium, the interface will adapt to a  
long--ranged deformation and  induce a capillary interaction energy 
$\propto \ln d$ between the colloids \cite{Kra00}, the
well--known {\em cheerios} effect \cite{Vel05}.
The occurence of such a long--ranged interaction 
is tied to the occurence of a net force on the system ``colloids + interface"
\cite{For04,Oet05,Oet06}.

\begin{figure}
 \begin{center}
  \epsfig{file=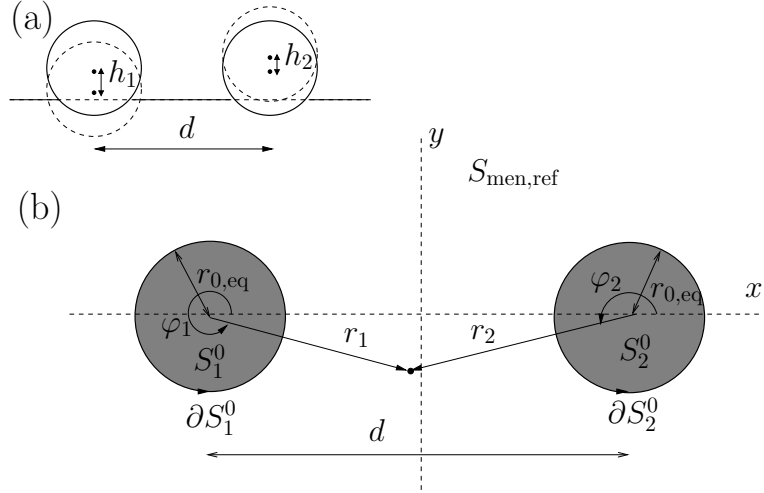, width=10cm}
 \end{center}
 \caption{ (a) Side view of two colloids at distance $d$ trapped at an 
interface, with $h_i$ $(i=1,2)$ denoting the relative center position
of the fluctuating colloid $i$ (dashed circles) with respect to
the colloid $i$ in equilibrium (full circles). (b)
Top view on the equilibrium interface, spanning the $x-y$--plane.
The colloids in equilibrium occupy the circular areas $S_i^0$, and
their boundaries $\partial S_i^0$ are the 
equilibrium (or reference) three--phase contact lines.
}
 \label{fig:twocolls}
\end{figure}

Here, we are interested in the occurence of an interface--mediated interaction
potential $V(d)$
in the force--free situation which are brought about by the fluctuating capillary
waves. To that end, we can apply the partition function analysis developed
in the previous section, extended to the case of two spherical ``obstacles" trapped
within the interface. We will focus on two scenarios:
\begin{itemize}
 \item[(A1)] No external force acts on the colloids, therefore the colloids 
   are free to fluctuate in the direction perpendicular
   to the interface. 
 \item[(A2)] The colloids are fixed at their equilibrium position by external 
   means. On average, there is no external force acting vertically on the
   colloids, although at a given instant of time some force is needed to 
   counteract the Brownian fluctuations of the colloids. In this case, 
   the effective potential $V(d)$ between the colloids is related to the
   pair correlation function $g(z_{\rm eq}, z_{\rm eq}, d)$ between the colloids    
   through $\beta V(d) = - \ln g(z_{\rm eq}, z_{\rm eq}, d)$. 
\end{itemize} 
In both scenarios, the interface and in particular the three--phase contact line
are free to fluctuate, subject to the energy penalty of the capillary wave and 
the boundary Hamiltonian lined out in the previous section.
In previous work \cite{Leh06,Leh07} (neglecting line tensions) we have 
established that the long--range
behavior of the effective potential $V(d)$ depends sensitively on the 
types of contact line fluctuations. For the case of a {\em pinned} contact line
on the colloid and the colloids fixed at their equilibrium position,
$V(d) \propto \ln\ln d$ whereas for the case (A1), $V(d) \propto d^{-8}$.
In the present work, we will show (i) that for case (A2), fixed colloids but 
unpinned contact line, the effective potential is still long--ranged,
$V(d) \propto \ln(1+\ln d)$ and (ii) that line tensions do not change the 
leading power in the long--range behavior of $V(d)$ but the instability for 
negative line tensions in the partition sum for a single colloid also occurs 
in the leading interaction  term for case (A1).

The effective potential $V(d)$ is obtained via the partition function of
the fluctuating capillary waves $\mc{Z}$ via
\bea
  V(d) = -\kbt \ln \mc{Z}(d)\;.
\eea 
The partition function $\mc Z \propto \int \mc{D}u\,\exp(-\beta\mc{H})$
contains a Hamiltonian which as before contains a 2d-``bulk" term
and a sum of boundary terms for each of the two colloids:
\bea
  \mc{H}[u(\vect x),z] &=& \left.\mc{H}_{\rm cw}[u(\vect x)]
    \right|_{\vect x \in S_{\rm men,ref}} +
   \sum_{i=1}^2 \left.\mc{H}^i_{\rm b}[u(\vect x),z]
   \right|_{\vect x \in \partial S^0_i}\;.
\eea
The functional form of $\mc{H}_{\rm cw}$ and the $\mc{H}^i_{\rm b}$ is given
by Eqs.~(\ref{eq:hcw}) and (\ref{eq:hb})--(\ref{eq:hb2}), respectively,
with due generalization of contact line multipoles for each colloid $i$,
$P_m\to P_{im}$, and of colloid height differences $h \to h_i=z_i-z_{\rm eq}$. 
In the equilibrium (reference) configuration, colloid $i$
intersects the interface plane in the circular area $S^0_i$, thus
the three--phase contact lines in the reference configuration are given by
$\partial S^0_i$ ($i=1,2$). Thus the 2d-``bulk" area over which the capillary waves 
fluctuate is given by 
$S_{\rm men,ref}=\mathbb{R}^2\setminus\bigcup_{i=1}^2 S^0_i$
(see also Fig.~\ref{fig:twocolls} for the geometric definitions).
Via the integration domain $S_{\rm men,ref}$ of $\mc{H}_{\rm cw}$,
 the total Hamiltonian and hence the partition function $\mc{Z}(d)$
of the system depends
on the distance $d$ of the colloid centers.

As before, the fluctuations over the contact lines 
$f_i=\sum_m \exp({\rm i}m\varphi_i)P_{im}$ 
are incorporated into the partition function
via $\delta$--function constraints:
\bea\label{eq:Z1}
\mathcal{Z}(d)&=&\mathcal{Z}_0^{-1} \int \mathcal{D}u\,
\exp \left(-\beta\mathcal{H}_{\rm cw}[u,d]  \right)\; \times \\
 \nonumber & & 
\prod_{\rm i=1}^2 
\int \mathcal{D}f_i
\prod_{{\bf x}_i \in \partial S^0_{i}}
\delta [u({\bf x}_{ i})-f_{ i}({\bf x}_{ i})]
\exp \left(-\beta \mathcal{H}^i_{{\rm b}}[f_i,h_i]\right)
\;.
\eea
The normalisation factor $\mc{Z}_{0}$
is chosen such that $\mc{Z}(d\to\infty)\to 1$. The difference between cases
(A1) and (A2) defined above shows up in the definition of the measure for
the contact line fluctuations:
\bea
 \mc{D} f_i &=& \left\{ \begin{array}{lr}
   dh_i\,dP_{i0}\prod_{m>0} d{\rm Re}P_{im}d{\rm Im}P_{i-m} & \quad \mbox{(A1)} \\
    \\
   dP_{i0}\prod_{m>0} d{\rm Re}P_{im}d{\rm Im}P_{i-m} & \quad \mbox{(A2)} 
 \end{array} \right. \;.
\eea
As seen above, in the unconstrained case (A1) an additional integral over
the colloid height variables is performed.
Both cases can be discussed conveniently by splitting the field $u$ of the local interface position into
a mean--field and a fluctuation part, $u=u_{\rm mf}+v$. The
mean--field part solves the Euler--Lagrange equation
$ (-\Delta+\lambda_c^{-2})\,u_{\rm mf}=0$ with the boundary condition
$u_{\rm mf}\,|_{ \partial S^0_{i}}=f_{ i} $. Consequently the fluctuation part
vanishes at the contact line:
$v\,|_{ \partial S_{ i}^0}=0$.
Then the partition function  $\mathcal{Z}=\mathcal{Z}_{\rm fluc}\mathcal{Z}_{\rm mf}$
factorises into a product of a fluctuation part
independent of
the boundary conditions
 and a  mean field part which depends on the fluctuating boundary conditions $f_i$
of the meniscus on the colloid surfaces:
\begin{eqnarray}\label{Zsep}
 \fl
 \mathcal{Z}_{\rm fluc} &=& \mathcal{Z}_0^{-1} \int \mathcal{D}v\,
\prod_{\rm i=1}^2 \prod_{{\bf x}_i \in \partial S^0_i}\delta ( v({\bf x}_i))
\exp \left(-\beta\mathcal{H}_{\rm cw}[v,d] \right)\;,  \nonumber \\ 
 \fl
 \mathcal{Z}_{\rm mf} &=& \prod_{\rm i=1}^2
\int \mathcal{D}f_i
\, \exp\left\{ -\frac{\beta\gamma }{2} \sum_{i}
\oint_{\partial S^0_i}d\ell_i f_i({\bf x}_i)\,( \partial_n u_{\rm mf}({\bf x}_i;d))\right\}
\exp \left(-\beta\mathcal{H}^i_{{\rm b}}[f_i,h_i] \right)
 \;.
\end{eqnarray}
The first exponential in $\mathcal{Z}_{\rm mf} $ stems from applying Gauss' theorem
to the energy associated with $u_{\rm mf}$. In this term
$ \partial_n u_{\rm mf}$ denotes the normal derivative of the mean--field
solution towards the interior of the circle $\partial S^0_i$, and $d\ell_i$ is
the infinitesimal line segment on $\partial S^0_i$ .

\begin{figure}
 \begin{center}
  \epsfig{file=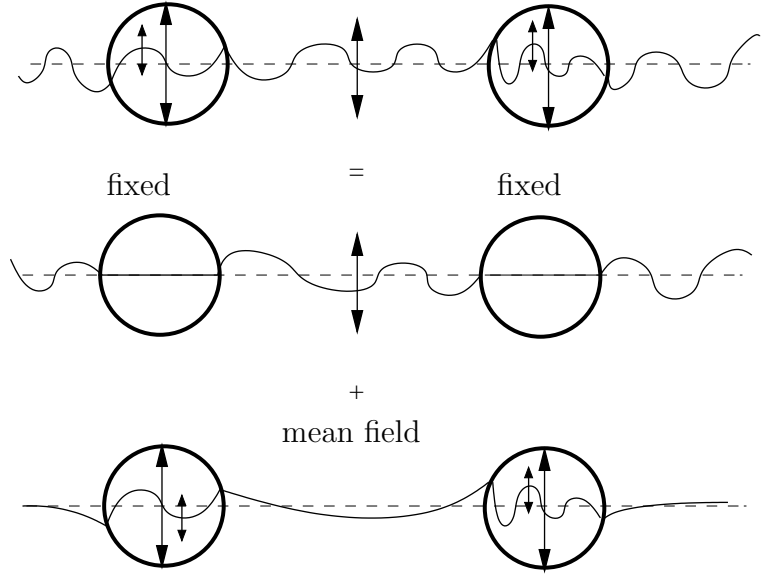, width=10cm}
 \end{center}
 \caption{ Pictorial representation of the separability of the total
 effective potential $V$ caused by the fluctuating capillary waves
 and the possibly fluctuating colloid
 into a fluctuation ($V_{\rm fluc}$) and a mean--field part ($V_{\rm men}$). 
 The fluctuation part (second picture) is obtained by summing over 
  all admissible capillary
 waves which are pinned at the colloid surface ($u=0$) and with the colloids
 themselves fixed. The mean--field part (third picture) is obtained
 by summing over the colloid height fluctuations (long arrows) and the 
 contact line fluctuations (short arrows) equipped with a Boltzmann
 factor incorporating the energy of the mean--field meniscus solution
 whose boundary condition is set by the momentary position of the contact 
 line. 
}
 \label{fig:flucsep}
\end{figure}

In this form, the partition function is amenable to analytical expansions for 
small and large distances $d$ between the colloids. The multiplicative 
separation of $\mc{Z}$ allows to define additive contributions 
to the effective potential:
$V=V_{\rm fluc}+V_{\rm mf}$ with 
$\beta V_{\rm fluc[mf]}=\ln\mc{Z}_{\rm fluc[mf]}$. The techniques to evaluate
the fluctuation and mean--field part to the effective potential have
been presented in detail in Ref.~\cite{Leh07}, and we {give a summary of} the main
results which are  necessary to discuss the influence of the line tension terms on
$V(d)$.

\subsection{Fluctuation part}
\label{sec:fluc}

The fluctuation part contributes equally
for both
 cases (A1) and (A2) introduced above.
The $\delta$-functions
in the fluctuation part of the partition function
can be removed by using their integral representation via
 auxiliary fields $\psi_i ({\bf x}_i)$ defined on the
interface boundaries $\partial S^0_i$ ~\cite{Li91}. This enables us to integrate out
the field $u$ leading to
\begin{eqnarray}
\label{eq:zfluc}
\fl
 \mathcal{Z}_{\rm fluc} =
\int \prod_{i=1}^2 \mc{D}\psi_i\,
\exp\left\{
-\frac{k_{\rm B}T}{2\gamma}\sum_{i,j=1}^2
\oint_{\partial S^0_i}d\ell_i \oint_{\partial S^0_j}d\ell_j\,
\psi_i ({\bf x}_i)\,G(|{\bf x}_i-{\bf x}_j|)\,\psi_j({\bf x}_j)\right\}\;.
\end{eqnarray}
We note in passing, that
 the fluctuation part in the form of Eq.~(\ref{eq:zfluc})
  resembles 2d screened electrostatics: it is the
partition function of a system
of fluctuating charge densities $\psi_i$ residing on the contact circles. For large
$d/r_0$ it can be calculated by utilizing the multipole expansion
\bea
   \psi_i(\varphi_i) = \sum_{m=-\infty}^\infty \psi_{im} \exp({\rm i}m\varphi_i)\;,
\eea
introducing the auxiliary multipole moments $\psi_{im}$ of order $m$ pertaining 
to colloid $i$. Using these it can be shown that in the limit
$\lambda_c \gg d \gg R$ the fluctuation part of the effective potential
$\beta V_{\rm fluc} = -\ln \mathcal{Z}_{\rm fluc}$  has the form
\bea
 \label{eq:flucexp}
  \beta V_{\rm fluc} (d) & = & {\rm const.} + a_0(d) 
  + \sum_{n=1}^\infty \frac{a_n}{d^{2n}}, \\
 \label{eq:a0}
    a_0(d) &=& \frac{1}{2} \ln \ln \frac{d}{r_{0,{\rm eq}}}  \;,
\eea 
where the $a_n$ for $n>0$ are numerical coefficients. Through the multipole analysis it is 
found that interaction terms $\propto \psi_{1m}\psi_{2m'}$ contribute terms
to $V_{\rm fluc}$ which are proportional to $d^{-2(m+m')}$.
The combination $\propto \psi_{10}\psi_{20}$ (fluctuating auxiliary monopoles)
gives rise to the leading term $a_0$ in $V_{\rm fluc}$. 

\subsection{Mean--field part}
The calculation of  $\mc{Z}_{\rm mf}$ (Eq.~(\ref{Zsep})) requires to determine
the solution of the differential equation 
\bea
 \label{eq:mf}
 (-\Delta+\lambda_c^{-2})\,u_{\rm mf}=0
\eea
with the boundary conditions at the fluctuating contact line and at infinity,
respectively: 
\bea
 \label{eq:mfb}
  \left.u_{\rm mf}({\bf x}_i)\right|_{{\bf x}_i \in \partial S^0_i}
    &=&f_i(\varphi_i) \\
  \left. u_{\rm mf}({\bf x})\right|_{|{\bf x}|\to \infty} &\to  &0 \;. 
\eea
We write the solution as a
 superposition  $u_{\rm mf} = u_1+u_2$
where $u_i= \sum_m  K_m ( r_i/\lambda_{ c})
A_{im} e^{{\rm i} m\varphi_i} $ is
the general mean--field solution in $\mathbb{R}^2\setminus  S^0_i$
{(see Fig.~\ref{fig:twocolls} for the geometric definitions)}.
The solution has to match to the boundary conditions
at both circles $\partial S^0_1$ and $\partial S^0_2$.
This can be achieved by  a projection of $u_2$ onto
the complete set of functions on $\partial S^0_1$, $\{ e^{{\rm i}m\varphi_1} \}$,
and vice versa.
Equating this expansion with the
contact line multipole expansion 
$f_i(\varphi_i)=\sum_i P_i\,\exp({\rm i}m\varphi_i)$
leads to a system of linear equations for the expansion coefficients
$\{A_{im}\}$. 
This system can be solved analytically within a systematic $1/d$ expansion or
numerically,
observing rapid
convergence. 
Owing to the linearity of Eq.~(\ref{eq:mf}), the  
mean field part of the partition function $\mc{Z}_{\rm mf}$
can be
written in a Gaussian form: 
\begin{equation}
\label{eq:zmf1}
\fl
\mc{Z}_{\rm mf} = \int \mc{D} f_i\; \exp\left ( - \beta \mc{H}[u_{\rm mf},d]
\right)\;
  \exp\left\{ -\pi \beta\gamma\left(1-\frac{\di \hat \tau}{\di \hat r_{0,{\rm eq}}^3}\right)\sum_i (P_{i0}-h_i)^2   \right\},
\end{equation}
where $\mc{H}[u_{\rm mf}]$ is a symmetric quadratic form in the vector of 
the contact line
multipole moments ${\bf {\hat f}}_i=(\dots, P_{i-1},P_{i0},P_{i1},\dots )$:
 \begin{eqnarray}\label{Hmean}
 \mc{H}[u_{\rm mf},d]
&=&
 \frac{\gamma}{2}
 \left(
 \begin{array}{c}
 {\bf\hat{f}}_1\\
 {\bf\hat{f}}_2
 \end{array}
 \right)
 ^{\rm T}
 \left(\begin{array}{cc}
 {\bf E}_{\rm 1\,self} & {\bf E}_{\rm int}(d) \\
 {\bf E}_{\rm int}(d) & {\bf E}_{\rm 2\, self}
 \end{array}\right)
 \left(
 \begin{array}{c}
 {\bf \hat{f}}_1\\
 {\bf \hat{f}}_2
 \end{array}
 \right)\;.
 \end{eqnarray}
Using this form, it can be shown that the mean--field part 
$V_{\rm mf}$ of the effective potential has a similar expansion to
the one of the fluctuation part $V_{\rm fluc}$ (Eq.~(\ref{eq:flucexp}))
in the limit $\lambda_c \gg d \gg R$:
\bea
  \beta V_{\rm mf} (d) &=& {\rm const.} + b_0(d) + 
  \sum_{n=1}^\infty \frac{b_n}{d^{2n}}\;. 
\eea
Also similar to the analysis of $V_{\rm fluc}$, interaction terms 
$\propto P_{1m}P_{2m'}$ contribute terms
to $V_{\rm mf}$ which are proportional to $d^{-2(m+m')}$.
The fluctuating contact line monopoles and the possibly (in case (A1)) 
fluctuating colloid heights $h_i$
gives rise to the leading term $b_0$ in $V_{\rm mf}$. 
The form of $b_0(d)$ and the values of the numerical coefficients
$b_n$ depend on the cases (A1) and (A2) introduced above.  
\begin{itemize}
 \item[(A1)] Here, for the freely fluctuating colloid, the integration measure
was given by $\mc{D}f_i =dh_id{\bf {\hat f}}_i$. Upon change of variables
$h_i \to h_i-P_{i0}$, it is seen from Eq.~(\ref{eq:zmf1}) that the 
$d$--dependent part of $\mc{Z}_{\rm mf}$ is given by $\det {\bf E}$.   
For the case of vanishing line tension, the properties of $\det {\bf E}$
were discussed in detail in Ref.~\cite{Leh07}. In particular it turns out that
the four leading terms in the expansion of $V_{\rm mf}$ and $V_{\rm fluc}$
cancel each other ($b_m=-a_m$ for $m=0,1,2,3$) and the total effective potential
is given to leading order by  
\bea
  \beta V(d) &\approx & \frac{a_4+b_4}{d^8} \;. 
\eea
This corresponds to a quadrupole--quadrupole interaction according to the
power counting in terms of the multipoles of either the fluctuating auxiliary
charge densities $\psi_{im}$ (for $V_{\rm fluc}$) or the fluctuating contact line
$P_{im}$ (for $V_{\rm mf}$). Hence, a nonvanishing line tension does not change
the leading power in $V(d)$ since the line tension contributions 
(via $\mc{H}_{b,2}$, see Eq.~(\ref{eq:hb2})) to $\det {\bf E}$ are nonzero only for
contact line multipoles higher than dipoles.\footnote{The boundary Hamiltonian
contains also a line tension contribution for the monopole terms 
$\propto (P_{i0}-h_i)^2$ (see Eq.~(\ref{eq:hb2})). 
Upon integration over the colloid height $h_i$,
the line tension dependence here is absorbed in a $d$--independent 
multiplicative factor in $\mc{Z}_{\rm mf}$, see Eq.~(\ref{eq:zmf1}).}
Therefore, including the line tension we find for the total effective potential
\bea
 \label{eq:a1}
  \beta V(d) & =& {\rm const.}  -\frac{1-\frac{\di 3\, \hat \tau}
   {\di \hat r_{0,{\rm eq}}^3}\,
\left(2-\frac{\di 3 \hat \tau}{\di \hat r_{0,{\rm eq}}^3} \right) }
{\left(1+\frac{\di \hat \tau}{\di \hat r_{0,{\rm eq}}^3}\right)^2}\;
 \frac{1}{d^8}
\eea
The dependence of the effective potential on $\tau$ is depicted in 
Fig.~\ref{fig:a1line}.
For positive line tensions, $\beta V(d)$ is always asymptotically attractive,
save for the value $\hat \tau/r_{0,{\rm eq}}^3 = 1/3$ where the coefficient
of the leading $d^{-8}$ term vanishes, and the effective potential becomes even
shorter ranged. For negative line tensions, we encounter a divergence of this
coefficient for $\hat \tau/r_{0,{\rm eq}}^3 \to -1$. This 
divergence is related to the instability in the one--colloid partition function
at the interface already discussed following Eq.~(\ref{eq:instability}). 
\begin{figure}
 \begin{center}
  \epsfig{file=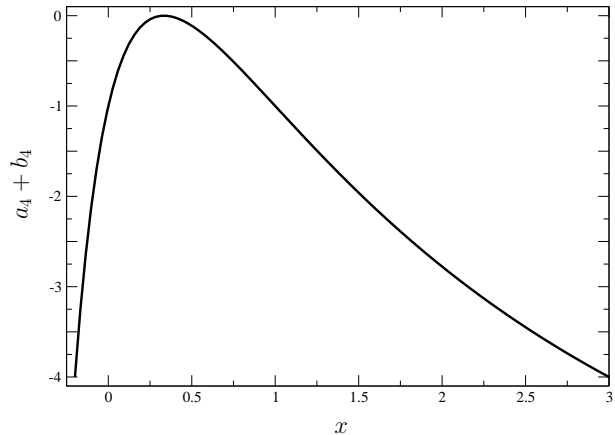, width=8cm}
 \end{center}
 \caption{For case (A1), the freely fluctuating colloid, the dependence of the 
coefficient of the leading $d^{-8}$ term ($a_4+b_4$) 
in the effective potential $V(d)$
on the line tension is shown. The dependence on $\tau$ enters through the
reduced variable $x= \hat \tau/r_{0,{\rm eq}}^3$ where $\hat \tau = \tau/(\gamma R)$
and $\hat r_{0,{\rm eq}}= r_{0,{\rm eq}}/R$.  }
 \label{fig:a1line}
\end{figure}
 \item[(A2)] This case implies fixing the colloids at their equilibrium 
positions $z_{\rm eq}$. Thus in the integration measure for the contact
line fluctuations the integration over the colloid height $h=z-z_{\rm eq}$ is 
absent, $\mc{D}f_i = d {\hat{\bf f}}_i$. According to Eq.~(\ref{eq:zmf1}),
the quadratic form ${\hat{\bf f}}^T {\bf E}\; {\hat{\bf f}}$ in the
Gaussian integral  is slightly changed by the second exponential on the
right hand side of Eq.~(\ref{eq:zmf1}). In particular, this leads to a changed
leading coefficient $b_0$ in the mean--field free energy
\bea
    b_0(d) = - \frac{1}{2}\, \ln \ln \frac{d}{r_{0,{\rm eq}}} + 
   \frac{1}{2}\, \ln \left( 1 + \left[ 1-\frac{\di  \hat \tau}
   {\di \hat r_{0,{\rm eq}}^3} \right] \ln \frac{d}{r_{0,{\rm eq}}} \right)\;.
\eea     
Therefore, for case (A2) the leading term in the total effective potential contains
a very long--ranged leading term of the form
\bea
 \label{eq:a2}
  \beta V(d) &\approx & {\rm const.} +
   \frac{1}{2}\, \ln \left( 1 + \left[ 1-\frac{\di  \hat \tau}
   {\di \hat r_{0,{\rm eq}}^3} \right] \ln \frac{d}{r_{0,{\rm eq}}} \right)\;,
\eea
which for $d \gg r_{0,{\rm eq}}$ slowly approaches the asymptotic
form found for $V_{\rm fluc}$ (Eq.~(\ref{eq:a0})). Thus for fixed colloid position
in the interface their pair correlation function 
$g(z_{\rm eq},z_{\rm eq},d)=-\exp(-\beta V(d))\propto - \ln(d/r_{0,{\rm eq}})$ 
contains a long--ranged piece dominated by the fluctuating ``bulk" capillary waves
only. This has also been found in a study treating the colloids as point 
particles \cite{Kai05} and in an analytical study of the pair correlation function
in phase--separating  2d and 3d lattice models \cite{Mac07}, and the asymptotics
of the colloid pair correlation function is the same as exhibited by the
{\em fluid} pair correlation function in the interface region \cite{Wer76}.
However, treating the finite size of the colloids correctly leads to sizeable
corrections in the asymptotics of the effective pair potential (Eq.~(\ref{eq:a2})) and
gives a nonzero $V(d)$ in the physically relevant case (A1) (it is zero in
the limit of point colloids). 
\end{itemize}

\section{Conclusion}

In this paper, we have studied the influence of capillary waves on the stability
and interactions of colloids (with radius $R$) trapped at a fluid interface
with surface tension $\gamma$, with particular
attention to the effects of a line tension $\tau$. 
Quite often, the stability of colloids
at a fluid interface with respect to vertical displacements $h$ from their
equilibrium position is discussed using a rigid interface model.
This gives for negligible line tensions and partially wetting colloids a steep
potential well with spring constant $k=2\pi\gamma$.
A finite line tension changes the spring constant by a term $\propto \tau/R$
and may induce metastability for the trapped colloids for certain positive
values of $\tau$. Within a perturbative model we have 
found that the potential well is considerably broadened by capillary waves
(qualitatively, $k \to k/\ln(\lambda_c/R)$ where $\lambda_c$ is the capillary
length in the interface system). This suggests also a reduction of the 
metastability barriers in case of positive line tensions, although
calculations beyond our perturbative model (quadratic in the fluctuations)
are needed for conclusive results.  

Capillary waves also induce effective interactions between two
colloids which are of Casimir type. For freely fluctuating colloids a
power-law dependence of the effective potential $V(d)$ in the intercolloidal distance
$d$ is obtained, $V \propto d^{-8}$. A finite line tension does not change
this power--law dependence, save for a specific positive value of $\tau$
where the corresponding coefficient vanishes and $V(d)$ becomes even 
shorter--ranged, decaying at least $\propto d^{-10}$.
Negative line tensions increase the amplitude
of $V(d)$. For colloids fixed in the interface, the effective potential is
equivalent to the potential of mean force between them and acquires a 
long--ranged component $V(d) \propto \ln(1+A\ln d)$ where $A$ is a line--tension
dependent coefficient (see Eq.~(\ref{eq:a2})). For $d \gg R$,
our results contain as a special case the long--ranged potential of mean force 
already discussed  for pointlike colloids within an fluctuating interface.     

In previous work \cite{Leh06,Leh07} we have discussed the strong 
attractive component in the fluctuation force which occurs for 
small separations between the colloid. This strong attraction is independent of
the surface properties and also of the line tension and can be understood 
from the capillary wave partition sum with stricht Dirichlet boundary conditions
on the colloid surface (see Subsec.~\ref{sec:fluc}). Both short--ranged and 
long--ranged regimes of the effective fluctuation potential should be 
important for the
aggregation of nanocolloids at interfaces and compete with 
other effective interactions such as of electrostatic origin
\cite{Bre07,Oet08}. 

{\bf Acknowledgment:} M. O. thanks the organizers of CODEF II for their
invitation and the German Science Foundation for
financial support through the Collaborative Research Centre
SFB-TR6 ``Colloids in External Fields”, project section D6-NWG.

\begin{appendix}

\section{Derivation of the boundary Hamiltonian}
\label{app:boundary}
In this appendix we derive the boundary term
$\mc{H}_{\rm b}$ which describes free energy changes upon shifting the
contact line
 (cf. the result in  Eqs.~(\ref{eq:hb})--(\ref{eq:hb2}) of Sec.~\ref{sec:cw}).
According to Eqs.~(\ref{eq:hs})--(\ref{eq:hcw}) the boundary term is given by
\bea
 \mc{H}_{\rm b}=
\gamma_{\rm I}\Delta A_{\rm I}+\gamma_{\rm II}\Delta A_{\rm II}
+\gamma\Delta A_{\rm proj}
+\tau \Delta L
\eea
and contains contributions associated with the difference in the interface areas colloid/phase I and colloid/phase II
between the configuration $\{u,z\}$ and the reference configuration 
$\{u=0,z_{\rm eq}\}$
(given by $\Delta A_{\rm I}$ and $\Delta A_{\rm II}$, respectively) and the
difference of the three--phase contact line length between these
configurations (given by $\Delta L$). The term $\Delta A_{\rm proj}$
describes the change in area (with respect to the reference configuration) of
meniscus $u(\vect r)$ projected onto the plane $z=0$.

If the three phase contact line is slowly varying without overhangs,
the following geometric relation holds between its projection onto 
the plane $z=0$ (parametrized in polar coordinates by $r_0(\varphi)$) and the
contact line  $u_0=u(r_0(\varphi),\varphi)$ itself:
\bea\label{tpcph} 
r_0(\varphi)
&=&
\left[
R^2-[u(r_0(\varphi),\varphi)-z]^2
\right]^{1/2}
\\
&=&
\left[
r_{0,\rm eq}^2+2\,z_{\rm eq}\,\left[u(r_0(\varphi),\varphi)- h\right]
-(u(r_0(\varphi),\varphi)-h)^2
\right]^{1/2} \;.
\nonumber
\eea
In Eq.~(\ref{tpcph}), $r_{0,\rm eq}=\sqrt{R^2-z_{\rm eq}^2}$
is the radius of the circular reference (or equilibrium) contact line,
$h=z-z_{\rm eq}$ is the deviation of the 
colloid center from its equilibrium height
and $u(r_0(\varphi),\varphi)$ is the actual height of the contact line
parametrized in terms of the polar angle $\varphi$.

Because of
$A_{\rm II}=4\pi R^2-A_{\rm I}$,
$\Delta A_{\rm I}=-\Delta A_{\rm II}$ holds
for fluctuations of the colloid surface area in contact with fluid I and II, respectively.
Then, the 
associated 
changes of the free energy
can be written as
\bea\label{Hbapp1}
\fl
\gamma_{\rm I}\Delta A_{\rm I}
+\gamma_{\rm II}\Delta A_{\rm II}
&=&
\gamma \cos\theta_0
\int_{0}^{2\pi}d\varphi\int_{r_{0,\rm eq}}^{r_0(\varphi)}
dr\,
\frac{r}{\sqrt{1-r^2/R^2}}
\nonumber \\
\fl &=& 
-R\gamma \cos\theta_0
\,\int_{0}^{2\pi}d\varphi\,\left[
u(r_0(\varphi),\varphi)-h
\right]
 \\
\fl &\simeq&
-\frac{\gamma R \cos\theta_0}{2z_{\rm eq}}\int_0^{2\pi}d\varphi\,
\left[
f- h
\right]^2
-\frac{\gamma R \cos\theta_0}{2z_{\rm eq}}\int_0^{2\pi}d\varphi\,
\left[
r_0^2(\varphi)-r_{0,{\rm ref}}^2
\right]\;, \nonumber
\eea
where  we have applied Eq.~(\ref{tpcph}).
Following Ref.~\cite{Oet05}, 
in the last line we have 
approximated 
the actual height of the contact line by the meniscus height at the
reference contact circle $\partial S_{\rm men,ref}$, i.e.
$u(r_0(\varphi),\varphi)\approx u(r_{0,{\rm eq}},\varphi)\equiv f(\varphi)$.
Correction terms to this approximation are at least of third order in
$u$ and $f$~\cite{Oet05}. 

The free energy contribution associated with the change in projected meniscus area can be written as
\bea\label{Hbapp2}
\gamma \Delta A_{\rm proj}
&=&
\gamma
\int_0^{2\pi}d\varphi\int_{r_0(\varphi)}^{r_{0,\rm eq}}
dr\,r
=\frac{\gamma}{2}\int_0^{2\pi}d\varphi\,
\left[
r_{0,\rm eq }^2-r_0^2(\varphi)
\right] \;.
\eea
Combining Eqs.~(\ref{Hbapp1}) and~(\ref{Hbapp2}),
applying again Eq.~(\ref{tpcph}) and using relation~(\ref{eq:zeq})
for the equilibrium position of the colloid 
($R\cos \theta_0 /z_{\rm eq}=\tau/(\gamma r_{0,\rm eq})-1$)
we find
\bea\label{Hbapp3}
\fl
\gamma_{\rm I}\Delta A_{\rm I}
+\gamma_{\rm II}\Delta A_{\rm II}
+\gamma \Delta A_{\rm proj}
&=&
\frac{\gamma}{2}\int_0^{2\pi}
d\varphi\,
[f-h]^2
-
\frac{\tau z_{\rm eq}}{r_{0,\rm eq}}\int_0^{2\pi}
d\varphi\, [f-h]
\;.
\eea

The free energy contribution related to the 
length fluctuations of the contact line
is written as
\bea\label{Hlin}
\tau \Delta L
&=&
\tau\int_{0}^{2\pi} d\varphi\,
\left[
\sqrt{r_0(\varphi)^2
+[\partial_\varphi r_0(\varphi)]^2
+[\partial_\varphi u(r_0(\varphi),\varphi)]^2
} 
-r_{0,\rm eq } \right]
\nonumber\\
&\simeq&
\frac{\tau}{2r_{0,\rm eq}}
\int_{0}^{2\pi}\!\!\!\! d\varphi\,
\left[
-\frac{R^2}{r_{0,\rm eq}^2}\,(f- h)^2
+2 z_{\rm eq}
\,(f- h)
+\frac{R^2}{r_{0,\rm eq}^2}
\,(\partial_\varphi f)^2
\right]
\eea
Comparing Eqs.~(\ref{Hbapp3})-(\ref{Hlin}), we find
that the linear terms cancel out because the equilibrium
position $z_{\rm eq}$ and $r_{0,\rm eq}$ are determined by
Eq.~(\ref{eq:zeq}).
Inserting the decomposition of $f(\varphi)$ from Eq.~(\ref{eq:conlin})
and performing the integrals over $\varphi$
finally leads to the form
$\mc{H}_{\rm b} \simeq \mc{H}_{\rm b,1}+\mc{H}_{\rm b,2}$ 
given in Eqs.~(\ref{eq:hb})--(\ref{eq:hb2}) for the
the total boundary Hamiltonian,
where the two contributions read
\bea\label{Hbapp4}
\mc{H}_{\rm b,1}
&=&\frac{\pi\gamma}{2}\left[
2(P_{0}-h)^2
+4\sum_{m\ge 1}|P_{m}|^2
\right]\;,
\\
\label{Hbapp5}
\mc{H}_{\rm b,2}
&=&
\frac{\pi R^2\tau}{2r_{0,{\rm eq}}^3}
\left[
-2
\,
(P_{0}-h)^2
+
4
\sum_{m=1}^{\infty}
\left(
m^2-1
\right)
\,|P_{m}|^2
\right] \;,
\eea
and describe changes in colloid surface energy and in line energy, respectively,
upon shifting the three phase contact line.

\section{Derivation of the mean--field energy term in Eq.~(\ref{eq:umf})}
\label{app:umf}

Let $(r,\varphi)$ be polar coordinates in the equilibrium interface plane
$z=0$ where $r=0$ is the center of the circle enclosed by the reference
contact line.
The solution to the mean--field equation 
$ (-\Delta+\lambda_c^{-2})\,u_{\rm mf}=0$
with the boundary condition $u_{\rm mf}(r_{0,{\rm eq}},\varphi)=
f(\varphi) = \sum_m P_m \exp({\rm i}m\varphi)$ is given by
\bea
  u_{\rm mf}(r,\varphi) =\sum_{m=-\infty}^\infty P_m\, \exp({\rm i}m\varphi)\, 
   \frac{K_{|m|}\left(\frac{\di r}{\di \lambda_c}\right)}
   {K_{|m|}\left(\frac{\di r_{0,{\rm eq}}}{\di \lambda_c}\right)}\;,
\eea
where $K_m$ is the modified Bessel function of the second kind and order
$m$. For nanocolloids, $\lambda_c \gg R$, so that one can use the 
approximation
\bea
   K_m(x) \approx \left\{ 
   \begin{array}{cr} -\ln(x/C) \qquad (C\simeq 1.12) & (m=0) \\
                   (2m-2)!!\,x^{-m}  & (m>0)
   \end{array} \right.
\eea
which is valid for $x \ll 1$. Thus we find
(with $f' = df/dr$)
\bea
 \fl
  \int_0^{2\pi} d\varphi \left(- u_{\rm mf}'(r_{0,{\rm eq}},\varphi)\,
  u_{\rm mf}(r_{0,{\rm eq}},\varphi)\right) \approx \frac{2\pi}{r_{0,{\rm eq}}}
  \left(  \frac{P_0^2}{\ln (C\lambda_c/r_{0,{\rm eq}})} + 
   2\sum_{m= 1}^\infty m|P_m|^2 \right)
\eea
which immediately leads to Eq.~(\ref{eq:umf}).

\end{appendix}


\section*{References}

\begin{thebibliography}{99}

\bibitem{Bor01} Bordag M, Mohideen U and Mostepanenko V M 2001
                {\it Phys. Rep.} {\bf 353} 1
\bibitem{Jaf04} Jaffe R L and Scardicchio A 2004 \PRL {\bf 92} 070402
\bibitem{Bue04} B{\"u}scher R and Emig T 2004 \PR A {\bf 69} 062101
\bibitem{Kre94} Krech M 1994 {\it The Casimir Effect in Critical Systems}
                (Singapore: World Scientific) 
\bibitem{Bec07} Hertlein C, Helden L, Gambassi A, Dietrich S and Bechinger C
                2008 {\it Nature} {\bf 451} 172
\bibitem{Kar99} Kardar M and Golestanian R 1999 
                {\it Rev. Mod. Phys.} {\bf 71} 1233
\bibitem{Kai05} Kaidi H, Bickel T and Benhamou M 2005 {\it EPL} {\bf 69} 15
\bibitem{Aar04} Aarts D G A L , Schmidt M and Lekkerkerker H N W 2004
                {\it Science} {\bf 304}  847
\bibitem{Gol96} Golestanian R, Goulian M and Kardar M 1996 \PR E {\bf 54} 6725
\bibitem{Emi08} Emig T, Graham N, Jaffe R L and Kardar M 2008 
                \PR D {\bf 77} 025005  
\bibitem{Gol00} Golestanian R 2000 \PR E {\bf 62} 5242
\bibitem{Leh06} Lehle H, Oettel M and Dietrich S 2006 {\it EPL} {\bf 75} 174
\bibitem{Leh07} Lehle H and Oettel M 2007 \PR E  {\bf 75} 011602
\bibitem{Pie80} Pieranski P 1980 \PRL {\bf 45} 569
\bibitem{Ave96} Aveyard R and Clint J H 1996 
                {\it J. Chem. Soc. Farad. Trans.} {\bf 92} 85                 
\bibitem{Bre99} Bresme F and Quirke N 1999 \JCP {\bf 110} 3536
\bibitem{Bre99a} Bresme F and Quirke N 1999 
                 {\it Phys. Chem. Chem. Phys.} {\bf 1} 2149
\bibitem{Gib61} Gibbs J W 1961 
                {\it The Scientific Papers of J. Willard Gibbs} 
                vol~1 (Ox Bow Press Connecticut) p 288
\bibitem{Row02} Rowlinson J S and Widom B 2002 
                {\it Molecular Theory of Capillarity} (New York: Dover)
\bibitem{Ami04} Amirfazli A and Neumann A W 2004 
                {\it Adv. Coll. Int. Sci.} {\bf 110} 121
\bibitem{Dre96} Drelich J 1996 {\it Colloids Surf. A} {\bf 116} 43
\bibitem{Schim07} Schimmele L, Napi\'orkowski M and Dietrich S 2007 
                  \JCP  {\bf 127} 164715 
\bibitem{Par07} 
  Park B J, Pantina J P, Furst E, Oettel M, Reynaert S and Vermant J
  2007 {\it Langmuir}  {\bf 24} 1686
\bibitem{Mec99} Mecke K and Dietrich S 1999 \PR E {\bf 59} 6766
\bibitem{Dai00} Fradin C, Braslau A, Luzet D, Smilgies D, Alba M, Boudet N, 
                Mecke K and Daillant J 2000  {\it Nature} {\bf 403} 871
\bibitem{Mil02} Milchev A and Binder K 2002 {\it EPL} {\bf 59} 81
\bibitem{Dai03} Mora S, Daillant J, Mecke K, Luzet D, Braslau A, Alba M 
                and Struth B 2003 \PRL {\bf 90} 216101
\bibitem{Tol04} Madsen A, Seydel T, Sprung M, Gutt C, Tolan M and 
                Gr\"ubel G 2004 \PRL {\bf 91} 096104
\bibitem{Vin05} Vink R, Horbach J and Binder K 2005  \JCP {\bf 122} 134905
\bibitem{Tar07} Tarazona P, Checa R and Chac\'on E 2007 \PRL {\bf 99} 196101
\bibitem{Oet05} Oettel M, Dominguez A and Dietrich S  2005 \PR E {\bf 71} 051401
\bibitem{Buf65} Buff F P, Lovett A and Stillinger F H 1965 \PRL {\bf 15} 621 
\bibitem{Klei_book} Kleinert H 1990 {\it Path integrals} 
                (Singapore: World Scientific) chapter 5 
\bibitem{Kra00} Kralchevsky P A and Nagayama K 2000 
                {\it Adv. Coll. Interface Sci.} {\bf 85} 145 
\bibitem{Vel05} Vella D and Mahadevan L 2005 {\it Am. J. Phys.} {\bf 73} 817 
\bibitem{For04} Foret L and W{\"u}rger A 2004 \PRL {\bf 92} 058302
\bibitem{Oet06} Oettel M, Dominguez A and Dietrich S 2006 
                {\it Langmuir} {\bf 22} 846 
\bibitem{Li91}  Li H and Kardar M 1991 \PRL {\bf 67} 3275
\bibitem{Mac07} Abraham D B, Essler F H and Maciolek A 2007 \PRL {\bf 98} 170602
\bibitem{Wer76} Wertheim M S 1976 \JCP {\bf 65} 2377 
\bibitem{Bre07} Bresme F and Oettel M 2007 \JPCM {\bf 19} 413101
\bibitem{Oet08} Oettel M and Dietrich S 2008 {\it Langmuir} {\bf 24} 1425 









\end{thebibliography}
\end{document}